\documentclass[aps,prd,twocolumn,amsmath,amssymb,groupedaddress]{revtex4-1}

\usepackage[utf8]{inputenc}
\usepackage[dvips]{graphicx}
\usepackage{psfrag}
\usepackage{subfigure}

\renewcommand{\vec}[1]{\ensuremath{\boldsymbol{#1}}}

\usepackage{xspace}
\newcommand*{\eg}{e.\,g.\@\xspace}
\newcommand*{\ie}{i.\,e.\@\xspace}
\newcommand*{\cf}{cf.\@\xspace}
\newcommand*{\eq}[1]{Eq.~(\ref{eq:#1})}
\newcommand*{\fig}[1]{Fig.~\ref{fig:#1}}
\renewcommand*{\sec}[1]{Sec.~\ref{sec:#1}}

\usepackage[ps2pdf=true]{hyperref}
\hypersetup{pdftitle=Neutrino Lump Fluid in Growing Neutrino
Quintessence}

\begin{document}


\title{Neutrino Lump Fluid in Growing Neutrino Quintessence}

\author{Youness Ayaita}
\email[]{y.ayaita@thphys.uni-heidelberg.de}
\author{Maik Weber}
\author{Christof Wetterich}
\affiliation{Institut für Theoretische Physik, Universität
Heidelberg\\ Philosophenweg 16, D--69120 Heidelberg, Germany}

\date{\today}

\begin{abstract}
	Growing neutrino quintessence addresses the {\it why now} problem
	of dark energy by assuming that the neutrinos are coupled to the
	dark energy scalar field. The coupling mediates an attractive
	force between the neutrinos leading to the formation of large
	neutrino lumps. This work proposes an effective, simplified
	description of the subsequent cosmological dynamics. We treat
	neutrino lumps as effective particles and investigate their
	properties and mutual interactions. The neutrino lump fluid
	behaves as cold dark matter coupled to dark energy. The methods
	developed here may find wider applications for fluids of composite
	objects.
\end{abstract}


\maketitle


\section{Introduction}
\label{sec:introduction}

The observed accelerated expansion of the Universe can be described by
a dark energy component \citep{Riess98, Perlmutter98}. Its energy
density dominates that of matter at present, while it constituted a
very small fraction of the energy budget in earlier stages of the
cosmic evolution \cite{Doran07, Reichardt11}. This ``why now''
problem has motivated the idea of dark energy being dynamically
coupled to other cosmological species. It has been proposed that a
dependence of the neutrino mass on the dark energy scalar field, the
cosmon, may naturally trigger the onset of accelerated expansion in
recent times \citep{Amendola07, Wetterich07}. The background
evolution of the resulting cosmological model, growing neutrino
quintessence, is similar to the concordance model with a cosmological
constant $\Lambda$.

\begin{figure}[htb!]
	\begin{center}
		\includegraphics[width=0.35\textwidth]{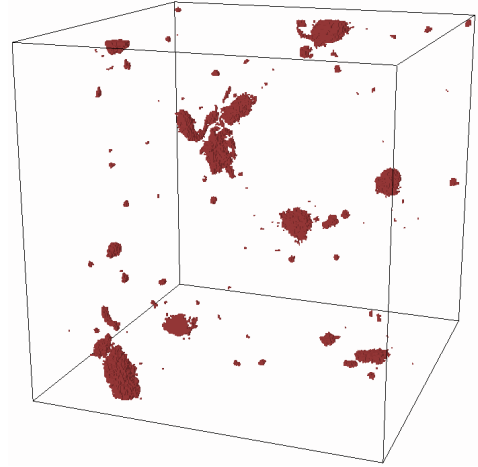}
	\end{center}
	\caption{Neutrino structures in a simulation box of comoving size
	$L = 600\, h^{-1}$Mpc at redshift $z = 1$. Shown are regions with
	a neutrino number density contrast above $5$ \citep{Ayaita11}.}
	\label{fig:lumps}
\end{figure}

Since the energy density in neutrinos is small, the cosmon-mediated
attraction between neutrinos has to be substantially stronger than the
gravitational one in order to be effective. This results in a fast
formation of neutrino lumps of the size of clusters or larger at
redshift around one. The dynamics of the perturbations in the coupled
cosmon-neutrino fluid is complicated. In contrast to models of
uncoupled or weakly coupled dark energy, a mere analysis of the
background equations together with linear perturbation theory is
insufficient. Linear perturbation theory breaks down even at large
scales \citep{Mota08}, and the nonlinear evolution exerts significant
backreaction effects on the background evolution. This has lead to
the development of a specifically designed N-body based simulation
method, which accounts for local cosmon perturbations, relativistic
neutrino motion, and backreaction effects \citep{Ayaita11}. These
simulations are, so far, successful until $z \approx 1$, where a
collection of spherical neutrino structures has formed, \cf
\fig{lumps}.

Although it is numerically challenging to resolve the internal
dynamics of the neutrino lumps, these details may not be crucial for
the broad cosmological picture. In gravity, \eg, the detailed
evolution inside galaxies or clusters is not relevant for the
cosmological evolution. Once the neutrino lumps have formed, one would
like to use a picture of a pressureless fluid of neutrino lumps.

In contrast to the universal properties of gravity, where only the
total mass of a bound object matters, the understanding of neutrino
lumps needs more information. The mass of a lump with a given number
of neutrinos is still expected to depend on the local value of the
cosmon field $\hat \varphi$ averaged in a region around the lump.
This effective coupling of $\hat \varphi$ to the lumps induces an
effective attractive interaction between the lumps. Since the lumps
are highly nonlinear objects, the $\hat\varphi$-dependence of the mass
is sensitive to the total number of neutrinos in the lump and possibly
even to additional properties of the lump.

This work presents analytical and numerical studies of the properties
of the neutrino lumps. We indeed find an effective description. This
opens the possibility for an approximate and much simpler approach to
the understanding of the cosmological evolution of growing neutrino
quintessence for the period after the formation of the lumps.

The paper is organized as follows. We collect some basics of growing
neutrino quintessence in Sec.~\ref{sec:fundamentals} and motivate the
approach taken in this work. Section~\ref{sec:effective} describes the
effective cosmological dynamics in the presence of stable neutrino
lumps. Starting from the basic idea of approximating lumps as
particles, we eventually develop a simplified simulation scheme of
growing neutrino quintessence. Some more technical aspects required
for this scheme are postponed to \sec{energy}. The question of
stability of neutrino lumps is discussed in \sec{aspects}. We conclude
in \sec{conclusion}.

\section{Fundamentals and motivation}
\label{sec:fundamentals}

After briefly summarizing the basics of growing neutrino quintessence
in \sec{basics}, we explain the main idea of this work. We give
physical arguments why the neutrino lumps may be approximated as
nonrelativistic particles. This forms the basis of the effective
description of the cosmological dynamics presented in \sec{effective}.

\subsection{Basics of growing neutrino quintessence}
\label{sec:basics}

The cosmon-neutrino coupling is described by the energy-momentum
exchange
\begin{align}
	\nabla_\lambda T^{\mu\lambda}_{(\varphi)} &= + \beta T_{(\nu)}
	\partial^\mu \varphi, \label{eq:exchange1} \\
	\nabla_\lambda T^{\mu\lambda}_{(\nu)} &= - \beta T_{(\nu)}
	\partial^\mu \varphi \label{eq:exchange2},
\end{align}
where $\beta$ is a dimensionless coupling parameter and $T_{(\nu)}
\equiv T^{\mu\lambda}_{(\nu)}g_{\mu\lambda}$ is the trace of the neutrino
energy-momentum tensor. We work in units where $8 \pi G = 1$ and use
the metric convention $ds^2 = -(1 + 2 \Psi) dt^2 + a^2 (1 - 2 \Phi)
d{\vec x}^2$. This type of coupling corresponds to early proposals of
coupled quintessence \cite{Wetterich94, Amendola99}. On the particle
physics level, the coupling is realized as a dependence of the
(average) neutrino mass $m_\nu$ on the cosmon field
\citep{Wetterich07}:
\begin{equation}
	\beta = - \frac{d\ln m_\nu}{d\varphi}.
	\label{eq:beta}
\end{equation}
For simplicity, we consider the case of a constant coupling parameter
$\beta$ as used in, \eg, \citep{Mota08, Baldi11, Ayaita11}. Typical
values are of order $\beta \sim -10^2$.

When the cosmon rolls down its potential towards larger values, a
negative $\beta$ implies a growing neutrino mass. As long as the
neutrinos are highly relativistic ($w_\nu \approx 1/3$), the trace
$T_{(\nu)} = -\rho_\nu (1 - 3 w_\nu)$ is close to zero and hence the
coupling is small. This changes once the neutrinos become
nonrelativistic. The coupling then stops the further evolution of the
cosmon resulting in an effective cosmological constant. In this way,
the model addresses the ``why now'' problem of dark energy. As in
standard quintessence models \citep{Wetterich88, Ratra88}, the energy
density of the dark energy scalar field $\varphi$ decays similarly to
the other species during most of the cosmological evolution, thereby
alleviating the fine-tuning of the present amount of dark energy.

The energy-momentum exchange, Eqs.~(\ref{eq:exchange1}) and
(\ref{eq:exchange2}), implies \citep{Wintergerst09}, in the Newtonian
limit, an attractive force between the neutrinos of order
\begin{equation}
	|\vec F| \approx |\beta \vec \nabla \varphi| \approx 2 \beta^2
	|\vec F_\text{gravity}|.
	\label{eq:fifthforce}
\end{equation}
We shall see that the interaction between neutrino lumps is similar
but with an effective coupling weaker than $\beta$.

\subsection{Lumps as nonrelativistic particles}
\label{sec:lumps}

The simulations of growing neutrino quintessence have shown that,
after a phase of rapid neutrino clustering, almost all cosmic
neutrinos are bound in roughly spherical lumps, \cf \fig{lumps}.

Inside these lumps, the neutrinos have relativistic velocities
\citep{Baldi11, Ayaita11}. For the neutrino fluid alone, one thus
observes a large pressure such that a nonrelativistic treatment is not
applicable. This is reflected in the equation of state $w_\nu = p_\nu
/ \rho_\nu$, which reaches $w_\nu \approx 0.1$ at $z = 1$
\cite{Ayaita11}. Nevertheless, we argue that the lumps as static bound
objects behave as particles with vanishing internal pressure. The
pressure induced by the neutrino motions is cancelled by a
corresponding negative pressure of the local cosmon perturbations.
Furthermore, the peculiar velocities of the lumps are nonrelativistic.
This is similar to a gas of atoms at low velocities. Although the
electrons move at high velocities, their contribution to the pressure
is cancelled by a contribution from the electromagnetic field.

Whereas the total pressure of a lump vanishes, the contributions of
neutrinos and the cosmon perturbation do not cancel locally. The
neutrinos are rather concentrated and hence their pressure
contribution is restricted to a small radius. The cosmon perturbation,
in contrast, extends to larger distances, analogously to the
gravitational potential around a massive object. The cancellation thus
only refers to the integrated contributions at a sufficiently large
distance from the lump.

\begin{figure}[htb!]
	\begin{center}
		\psfrag{xlabel}[B][c][.8][0]{physical radius $ar$
		[$h^{-1}$Mpc]}
		\psfrag{ylabel}[B][c][.8][0]{$P(r)/M_l$}
		\includegraphics[width=0.45\textwidth]{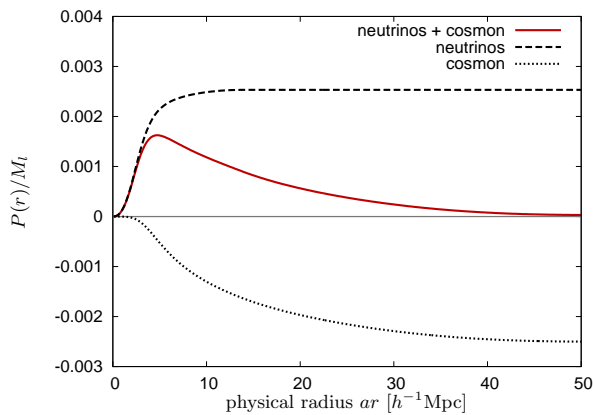}
	\end{center}
	\caption{Integrated pressure contributions $P_\nu$ (black dashed),
	$P_{\delta\varphi}$ (black dotted), and their sum (red solid),
	normalized by the lump mass $M_l$.}
	\label{fig:cancellation}
\end{figure}

In the following, we discuss this in more detail. Since gravity is
subdominant compared to the fifth force, \cf \eq{fifthforce}, it may
be neglected for a simple discussion. On general grounds, one can show
that a bound object has vanishing pressure if three conditions are
met:
\begin{enumerate}
	\item The object is described by a conserved energy-momentum
		tensor.
	\item The energy-momentum tensor vanishes outside a volume
		surrounding the object.
	\item The energy-momentum tensor is static.
\end{enumerate}
The argument is given in \sec{single}. For the purpose of
illustration, we have numerically simulated an exemplary spherical
neutrino lump satisfying these idealized conditions. The staticity of
the lump was realized by a hydrodynamic balance equation, \cf
\sec{hydrodynamic}. The neutrino pressure integrated to a comoving
radius $r$ from the center is given by a sum over particles $p$
\begin{equation}
	P_\nu(r) = \int_{0}^{r} 4\pi r^2 dr \sqrt{g^{(3)}}
	\frac{1}{3} T^i_{(\nu)\, i}
	= \sum_{p}^{} \frac{\gamma_p}{3} m_p \vec
	v_p^2,
	\label{}
\end{equation}
with the Lorentz factor $\gamma_p$ and the determinant of the spatial
metric $\sqrt{g^{(3)}} \approx a^3$. The contribution of the cosmon
perturbation is
\begin{align}
	P_{\delta\varphi}(r)
	&= -\int_{0}^{r} 4\pi r^2 dr \sqrt{g^{(3)}} \left[
	\frac{|\vec\nabla \delta\varphi|^2}{6a^2} +
	V'(\bar\varphi)\delta\varphi \right],
	\label{}
\end{align}
where we have subtracted the pressure induced by the background field
$\bar \varphi$. Figure~\ref{fig:cancellation} shows the cancellation
of the pressure contributions for large radii.
As already explained, the cosmon contribution is more extended than
the neutrino contribution.

In the cosmological context, the aforementioned conditions are met, at
best, approximately and realistic neutrino lumps will not be exactly
pressureless. We shall now discuss the three conditions. First, only
the total energy-momentum tensor of neutrinos, local cosmon
perturbation, and background cosmon is conserved. The background
field, however, cannot be attributed to the lump (otherwise, the
second condition would not be satisfied). The energy-momentum tensor
of the lump, defined to include the neutrinos and the local cosmon
perturbation, is thus not exactly conserved due to exchange between
the lump and the outside cosmon field. Finally, even for a virialized
lump with a fixed number of neutrinos, the energy-momentum tensor is
not static. Due to the time evolution of the outside cosmon field, the
mass of the neutrinos and therefore the mass of the lump changes.

One may argue that these effects are suppressed by the difference in
the relevant time scales for the dynamics of the lump and the
cosmological evolution. Indeed, the violations of staticity and
energy-momentum conservation are proportional to the time derivative
of the cosmon field averaged on length scales much larger than the
size of the lump. This is suppressed by the fact that the associated
time scale is large as compared to the dynamical time scale of the
lump. The effective description of growing neutrino quintessence
presented in the next section assumes that the pressure of neutrino
lumps approximately vanishes.

\section{Effective dynamics}
\label{sec:effective}

The approach of this section is to treat the neutrino lumps as
effective particles. We then merely have to characterize their mutual
interactions and their influence on the background as well as on the
gravitational potential. A numerical treatment of the internal
structure of the lumps will no longer be required.

In \sec{description}, we shall describe how lumps can be treated as
particles with an effective coupling. Section~\ref{sec:evolution}
derives the equation of motion for these effective particles and
explains how to calculate the relevant potentials: the large-scale
cosmon $\hat \varphi$ and the gravitational potential $\hat \Psi$.
Finally, we explain in \sec{simulation} how the results can be used to
construct the simplified simulation scheme for growing neutrino
quintessence.

\subsection{Description of lumps}
\label{sec:description}

Let us introduce a comoving length scale $\lambda$, which is larger
than the typical lump sizes but smaller than their typical distances
(the mean distance between neighboring lumps is of order $100
\,h^{-1}$Mpc). On scales larger than $\lambda$, a lump $l$ at comoving
coordinates $\vec x_l$ looks effectively point-shaped,
\begin{equation}
	T^{\mu\nu}_{l} \approx
	\frac{A^{\mu\nu}}{\sqrt{g^{(3)}}}\,\delta^{(3)}(\vec x - \vec
	x_l),
	\label{eq:tmunueff}
\end{equation}
the amplitude $A^{\mu\nu}$ being given by the integrated local
energy-momentum tensor of the lump,
\begin{equation}
	A^{\mu\nu} = \int d^3 y \, \sqrt{g^{(3)}}\,
	T^{\mu\nu}_\text{local} (\vec y).
	\label{eq:amunu}
\end{equation}
We will see in \sec{single}
that this indeed reduces to the standard one-particle case
\begin{equation}
	A^{\mu\nu} \approx \frac{M_l}{\gamma}\, u^\mu u^\nu,
	\label{eq:amunueff}
\end{equation}
where $M_l$ is the lump's rest mass (consisting of a neutrino and a
cosmon contribution) and $u^\mu$ is its four-velocity. The Lorentz
factor is defined as $\gamma = \sqrt{-g_{00}}\,u^0$. In the background
metric, we have $\gamma = u^0$. The result for $A^{\mu\nu}$ is a
consequence of the approximate pressure cancellation discussed in
\sec{lumps}.

The interactions between the lumps are mediated by the cosmon field
$\varphi$. Given that the distances between the lumps are greater than
$\lambda$, it suffices to consider the smoothed field (indicated by a
hat)
\begin{equation}
	\hat \varphi(\vec x) = \int d^3 y\,\sqrt{g^{(3)}}\, W_\lambda(\vec
	x - \vec y) \, \varphi(\vec y) \label{eq:smoothedfield}
\end{equation}
with a suitable window $W_\lambda$ of size $\lambda$.

Analogously to the fundamental coupling parameter $\beta$, \eq{beta},
we may define the effective coupling by
\begin{equation}
	\beta_{l} = -\frac{d\ln M_l}{d \hat \varphi}.
	\label{eq:betaeff}
\end{equation}
The effective coupling may depend on the scale $\lambda$ over which
the field is averaged. Whereas the fundamental coupling $\beta$
describes the dependence of the microscopic neutrino mass $m_\nu$ on
the local cosmon field $\varphi$, the effective coupling $\beta_{l}$
measures the mass dependence of the total lump mass $M_l$ on the
large-scale cosmon value $\hat \varphi$. As the fundamental parameter
$\beta$ quantifies the force between neutrinos, \cf \eq{fifthforce},
the effective parameter $\beta_{l}$ will determine the interactions
between lumps.

We next show quantitative results for the distribution of lumps and
the effective couplings at $z=1$. For this purpose, we have performed
$10$ simulation runs with the method and the parameters of
Ref.~\citep{Ayaita11}: fundamental coupling $\beta = -52$, box size $L
= 600\, h^{-1}$Mpc, but with reduced resolution $N_\text{cells} =
128^3$. The positions of the lumps have been identified as local
maxima of the neutrino density field (\cf DENMAX halo finding
\cite{Gelb92}). A glance at \fig{lumps} shows that there is not much
ambiguity in identifying lumps.

Once a stable lump has formed, the number of bound neutrinos is
approximately fixed (neglecting merging processes). It is thus natural
to characterize different lumps by their amount of neutrinos.

We measure the effective couplings $\beta_l$ and the lump masses
$M_l$. The latter include a (dominant) neutrino contribution
$M^{(\nu)}_l$ and a somewhat smaller cosmon part $M^{(\varphi)}_l$.
Integration over the comoving lump volume $V_l$ yields
\begin{align} 
	\gamma_l\,M^{(\nu)}_l &= \int_{V_l}^{} d^3 x\, \sqrt{g^{(3)}}\, \rho_\nu
	\approx
	\sum_{\text{particles }p}^{} \gamma_p m_{\nu,p}, \label{eq:mnul}\\
	\gamma_l\,M^{(\varphi)}_l &= \int_{V_l}^{} d^3 x\,
	\sqrt{g^{(3)}}\,(\rho_\varphi - \rho_{\hat \varphi}),
	\label{eq:mphil}
\end{align}
with $\rho_\varphi =\frac{{\dot \varphi}^2}{2} + \frac{|\vec \nabla
\varphi|^2}{2 a^2} + V(\varphi)$. The Lorentz factor $\gamma$ depends
on the velocities of the lumps or particles, respectively. The
smoothed field $\hat \varphi$ is considered external to the lump and
thus subtracted.

\begin{figure}[htb!]
	\begin{center}
		\psfrag{xlabel}[B][c][.8][0]{threshold $f$ }
		\psfrag{ylabel}[B][c][.8][0]{lump abundance $N(f_l> f)$ in $V_H$}
		\includegraphics[width=0.45\textwidth]{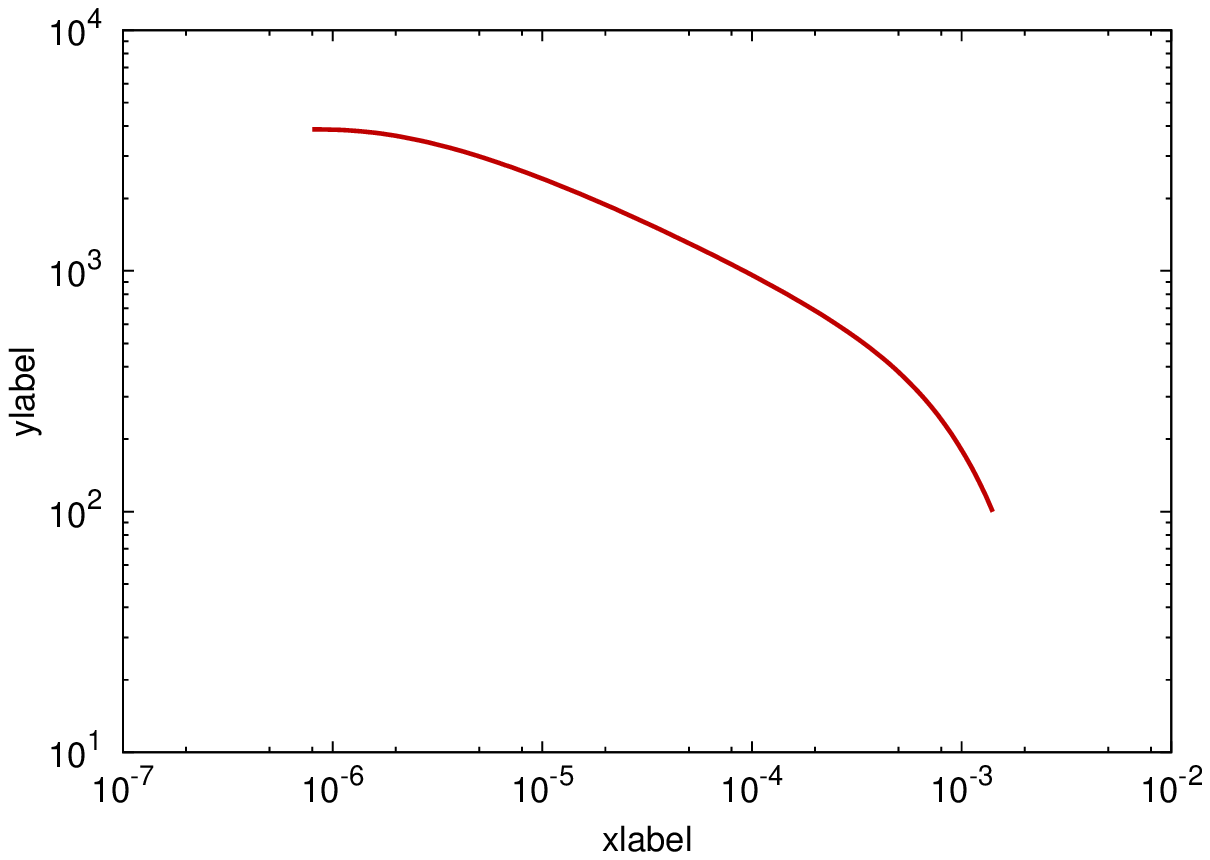}

		\psfrag{xlabel}[B][c][.8][0]{neutrino number fraction $f_l$ }
		\psfrag{ylabel}[B][c][.8][0]{effective coupling $\beta_l/\beta$}
		\includegraphics[width=0.45\textwidth]{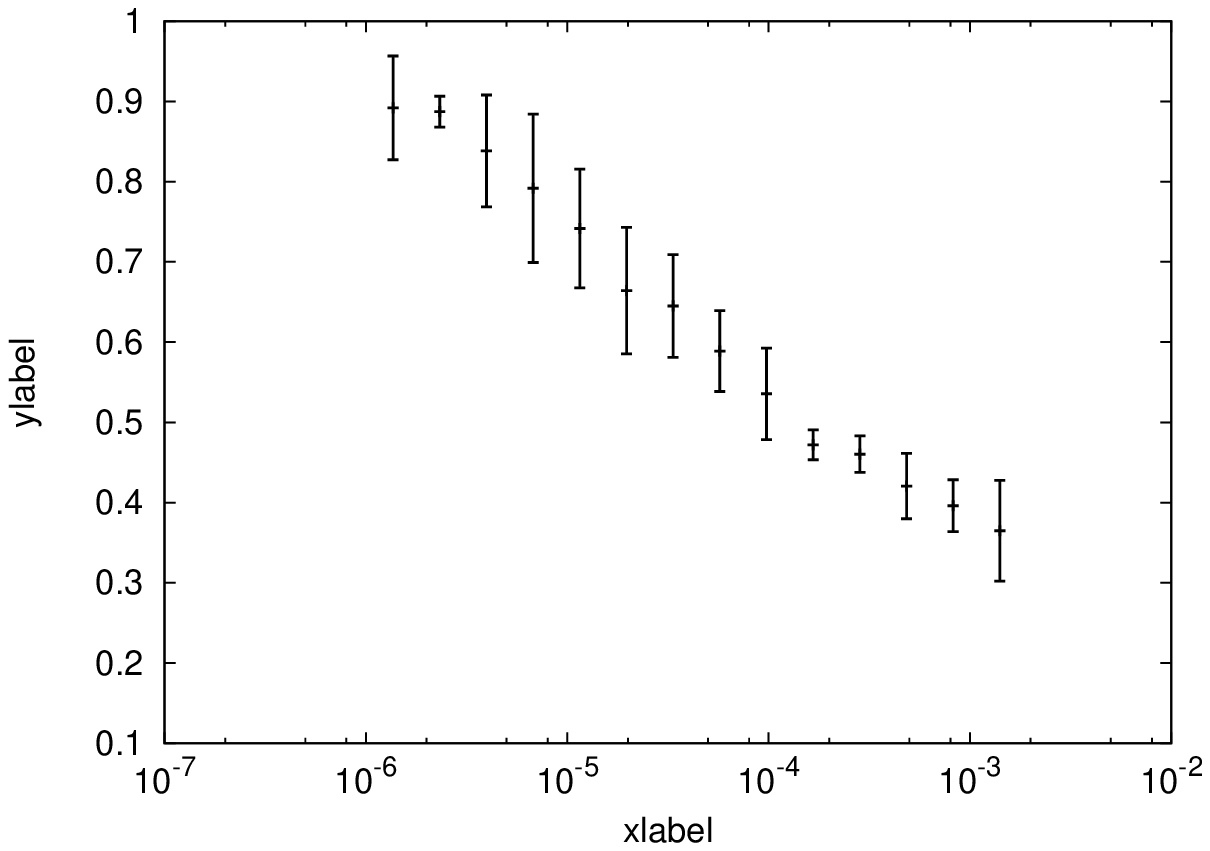}

		\psfrag{xlabel}[B][c][.8][0]{neutrino number fraction $f_l$ }
		\psfrag{ylabel}[B][c][.8][0]{lump mass $M_l/M_\odot$ }
		\includegraphics[width=0.45\textwidth]{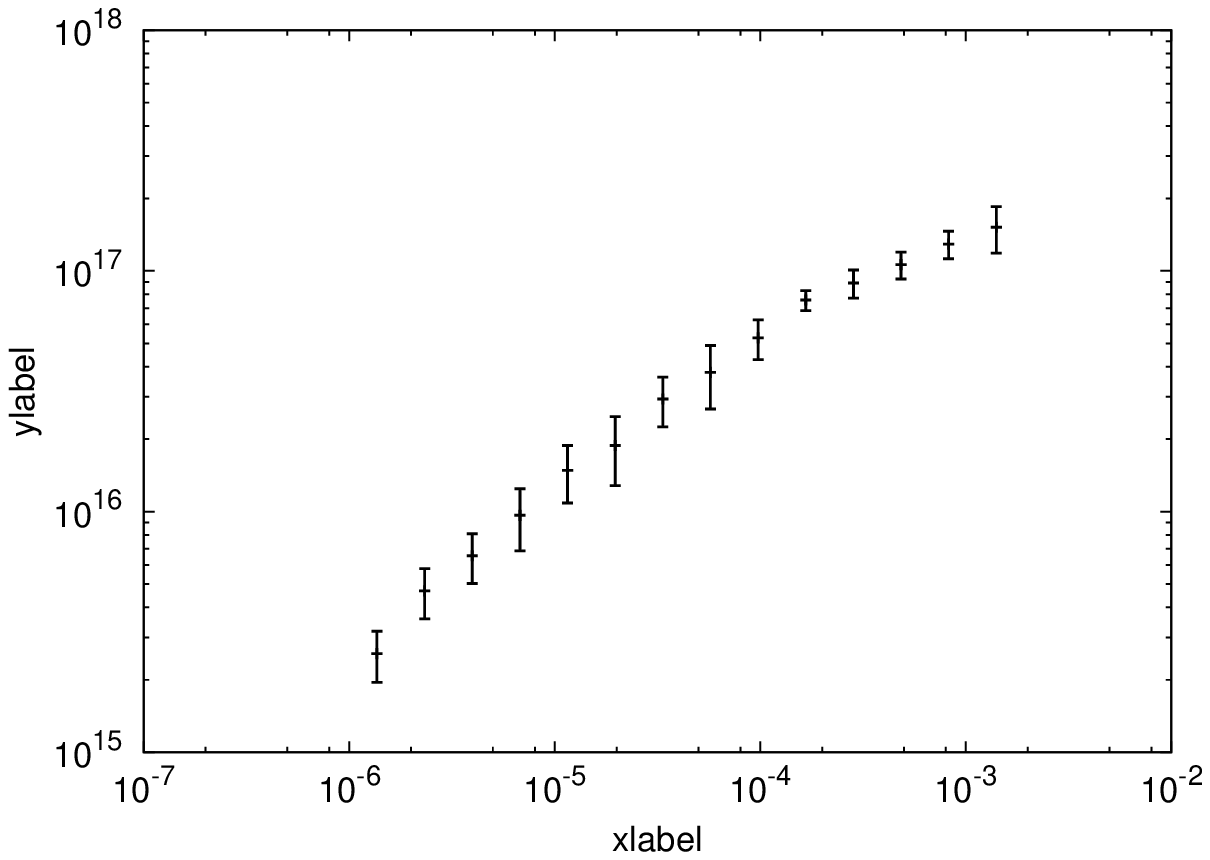}
	\end{center}
	\caption{Lump abundances, effective couplings, and lump masses as
	functions of the neutrino number fraction $f_l$ (number of
	neutrinos in the lump normalized to the number of neutrinos in the
	Hubble volume $V_H = H_0^{-3}$) at redshift $z = 1$. The error
	bars indicate the variance of lumps in the same bin.}
	\label{fig:massfunctionbetas}
\end{figure}

Figure~\ref{fig:massfunctionbetas} shows the abundance of lumps and
the distributions of $\beta_{l}$ and $M_l$.
The couplings $\beta_{l}$ are measured by numerical differentiation
according to \eq{betaeff}. The two lower figures show approximate
functional dependences on the neutrino amount with only relatively
small statistical fluctuations. The effective coupling is
systematically weaker than the fundamental coupling. This becomes more
pronounced with increasing neutrino number.

For the averaging scale $\lambda$, we have taken $\lambda = 30\,
h^{-1}$Mpc. This is clearly smaller than the typical lump distances
$\sim 100\, h^{-1}$Mpc but larger than the neutrino concentration of
the lumps. Concerning the cosmon field, there remains some ambiguity
since we attribute only the cosmon perturbations at scales smaller
than $\lambda$ to the lumps. If $\lambda$ is chosen larger, the
pressure cancellation and thus the particle approximation are better,
\cf \fig{cancellation}, but there may arise overlaps between spatially
close lumps.

\subsection{Evolution equations}
\label{sec:evolution}

Within our effective description, the equation of motion of a neutrino
lump is derived from the standard one-particle action
\begin{equation}
	S = \int d^4 x\, \sqrt{-g}\, T^{\mu\nu}_l g_{\mu\nu}
	= - \int d\tau\, M_l(\hat \varphi)
	\label{eq:oneparticleaction}
\end{equation}
with the proper time $\tau$ and the smoothed cosmon field $\hat
\varphi$ evaluated at the lump trajectory. Along the same lines as for
the single neutrino case \cite{Ayaita11}, we arrive at
\begin{equation}
	\frac{du^\mu}{d\tau} + \Gamma^\mu_{\rho\sigma} u^\rho u^\sigma
	= \beta_{l}\, \partial^\mu \hat \varphi
	+ \beta_{l}\,u^\lambda \partial_\lambda \hat \varphi\,
	u^\mu.
	\label{eq:eom}
\end{equation}
The left-hand side describes gravity (expansion and gravitational
potential), the right-hand side is due to the cosmon-neutrino
interaction. The (spatial) term $\beta_{l} \vec \nabla \hat \varphi$
is the cosmon-mediated fifth force analogous to Newtonian gravity, \cf
\eq{fifthforce}. The second contribution on the right-hand side
reflects momentum conservation: A lump is accelerated when it moves
towards a direction where it loses mass.

In order to use this effective equation of motion, we need to know the
smoothed cosmon field $\hat \varphi$, the gravitational potential
$\Psi$, and the background evolution. In the following, we shall
describe how this is achieved.

For the calculation of $\hat \varphi$, we recall the coupled
Klein-Gordon equation, separated in background and perturbation parts
\citep{Ayaita11},
\begin{align}
	\ddot {\bar \varphi} + 3 H \dot {\bar \varphi} + V'(\bar \varphi)
	&= -\beta {\bar T}_{(\nu)},
	\label{eq:kgbg}
	\\
	\Delta \delta\varphi - a^2 V''(\bar \varphi) \delta \varphi
	&= \beta\,a^2 \delta T_{(\nu)}.
	\label{eq:kgpert}
\end{align}
In \eq{kgpert} we have neglected the gravitational potential against
the cosmon perturbation. The second equation is similar to the
gravitational Poisson equation. A natural choice for the cosmon
potential $V$ is the exponential potential $V(\varphi) \propto
\exp(-\alpha \varphi)$ \citep{Wetterich08}.

Next, we will smooth the perturbation equation (\ref{eq:kgpert}). For
the left-hand side, it is straightforward to show by partial
integration that
\begin{align}
	\widehat{\Delta \delta\varphi} (\vec x)
	&= \int_{}^{} d^3 y\,
	\sqrt{g^{(3)}}\, W_\lambda(\vec x - \vec y) \Delta_{\vec y}
	\delta\varphi(\vec y)
	\nonumber \\
	&= \Delta_{\vec x} \int_{}^{}
	d^3 y\,\sqrt{g^{(3)}}\, W_\lambda (\vec x - \vec y)\,
	\delta\varphi(\vec y)
	\nonumber \\
	&= \Delta \delta\hat\varphi
	(\vec x), \label{}
\end{align}
up to surface terms and neglecting the metric perturbations,
$\sqrt{g^{(3)}} \approx a^3$. On the right-hand side, we write
$\widehat{\delta T_{(\nu)}} = \hat T_{(\nu)} - \bar T_{(\nu)}$
with the smoothed energy-momentum tensor of neutrinos 
\begin{equation}
	\hat T_{(\nu)}(\vec x) = \int_{}^{} d^3 y\,
	\sqrt{g^{(3)}}\, W_\lambda(\vec x - \vec y) T_{(\nu)}(\vec y).
	\label{}
\end{equation}
We next employ the relation (shown in \sec{smoothed})
\begin{equation}
	\beta {\hat T}_{(\nu)} \approx \sum_{\text{lumps }l}^{}
	\beta_{l} {\hat T}_{l}.
	\label{eq:smoothedsum}
\end{equation}
Here, the smoothed trace of the energy-momentum tensor of a lump $\hat T_l$ can be calculated
from the effective lump energy-momentum tensor,
Eqs.~(\ref{eq:tmunueff}) and (\ref{eq:amunueff}), $T_l = T^{\mu\nu}_l
g_{\mu\nu}$:
\begin{align}
	\hat T_l &= \int_{}^{} d^3 y\, \sqrt{g^{(3)}}\, W_\lambda(\vec x - \vec y)\,
	T_l(\vec y)
	\nonumber \\
	&= -\frac{M_l}{\gamma_l} W_\lambda(\vec x - \vec x_l).
	\label{}
\end{align}
With these results, the smoothed perturbation equation eventually
reads
\begin{align}
	\Delta \delta & \hat\varphi(\vec x) - a^2 V''(\bar \varphi)
	\delta\hat\varphi(\vec x)
	= \nonumber
	\\
	& -a^2 \sum_{\text{lumps }l}^{}
	\beta_{l} \frac{M_l}{\gamma_l} W_\lambda(\vec x - \vec x_l) -
	\beta a^2 {\bar T}_{(\nu)}.
	\label{eq:poissonsmooth}
\end{align}
Assuming that all
neutrinos are bound in lumps, one has
\begin{equation}
	\beta {\bar T}_{(\nu)} = -\frac{1}{V_\text{phys}}
	\sum_{\text{lumps }l}^{} \beta_{l} \frac{M_l}{\gamma_l}
	\label{eq:betatbar}
\end{equation}
in some cosmological volume $V_\text{phys}$.

On scales larger than $\lambda$, the
window $W_\lambda(\vec x - \vec x_l)$ in \eq{poissonsmooth} may be
replaced by a point $\propto \delta^{(3)}(\vec x- \vec x_l)$. For an
approximate solution of \eq{poissonsmooth} at distances larger than
$\lambda$ from the sources, we thus use a sum of Yukawa potentials,
\begin{equation}
	\delta\hat\varphi \approx \sum_{\text{lumps }l}^{} \left(
	\frac{\beta_{l}}{4\pi a}\,\frac{M_l/\gamma_l}{|\vec x - \vec
	x_l|}\,e^{-a m_\varphi|\vec x - \vec x_l|} +
	\delta\varphi_{\text{res},l} \right)
	\label{eq:phicalc}
\end{equation}
with the scalar mass $m_\varphi^2 \equiv V''(\bar \varphi)$.  The
residual term $\delta\varphi_{\text{res},l}(\vec x - \vec x_l)$ is
needed to cancel the background part $\propto \bar T_{(\nu)}$ on the
right-hand side and to ensure $\overline{\delta\varphi} = 0$ in a
simulation volume, similar to $\Psi_{\text{res},l}$ below.

If the lumps are moving rather slowly compared to the speed of light,
$\gamma_l \approx 1$, the two smoothed metric potentials are
equivalent, $\hat \Phi \approx \hat \Psi$. Numerically, this relation
is verified on large scales \cite{Ayaita11}. Then, we write for the
smoothed gravitational potential induced by lumps (with the same
approximations as for $\hat\varphi$)
\begin{equation}
	\Delta {\hat \Psi}(\vec x) \approx \frac{a^2}{2} \sum_{\text{lumps }l}^{}
	\left( M_l \frac{\delta^{(3)}(\vec x - \vec
	x_l)}{\sqrt{g^{(3)}}} -
	\frac{M_l}{V_\text{phys}} \right).
	\label{}
\end{equation}
The solution is (up to a constant)
\begin{equation}
	{\hat \Psi}(\vec x)
	= -\sum_{\text{lumps }l}^{} \left( \frac{1}{8
	\pi a}\,\frac{M_l}{|\vec x - \vec x_l|}
	+ \Psi_{\text{res},l}
	\right),
	\label{eq:psicalc}
\end{equation}
where the residual contribution can be given explicitly as
$\Psi_{\text{res},l} = M_l a^2 |\vec x - \vec x_l|^2/V_\text{phys}$.
The total gravitational potential also includes the matter-induced
potential which is calculated as usual. Taking into account
relativistic corrections would require the calculation of both
potentials, $\hat \Psi$ and $\hat \Phi$, \cf Ref.~\cite{Ayaita11}.

In order to have a full description of the cosmological dynamics, we
still need to describe the evolution of the cosmological background,
\ie the Hubble expansion $H$ and the background cosmon $\bar \varphi$.
The background evolution cannot be calculated without taking into
account the {\it backreaction} due to the perturbation evolution
\cite{Pettorino10, Ayaita11}. Instead, the background and the
perturbations have to be evolved simultaneously. In particular, one
averages first $\beta {\bar T}_{(\nu)}$ as in \eq{betatbar} and
inserts this into the background part of the Klein-Gordon equation
(\ref{eq:kgbg}). In every step, the perturbations enter the background
equations via $\beta_{l}$ and $M_l$.

\subsection{Simulation scheme}
\label{sec:simulation}

The methods developed in the previous sections allow for a
considerable simplification of the numerical treatment. Rather than
evolving a large number of N-body particles and the fields $\varphi$
and $\Psi$ on a grid, one now merely has to evolve a drastically
reduced set of differential equations. This becomes possible as soon
as a collection of stable neutrino lumps has formed (at about $z
\approx 1$). The preceding cosmological evolution has to be carried
out with the comprehensive simulation method of Ref.~\cite{Ayaita11}.
Its final state at $z \approx 1$ provides a distribution of lumps at
positions $\vec x_l$, with neutrino number factions $f_l$, rest masses
$M_l$, and effective couplings $\beta_{l}$. This is the starting point
for the simplified scheme.

Section~\ref{sec:evolution} collects a set of coupled differential
equations describing the cosmological evolution. These are the
equation of motion (\ref{eq:eom}), the background Klein-Gordon
equation (\ref{eq:kgbg}) with its right-hand side (\ref{eq:betatbar})
and the usual Friedmann equations. They involve the averaged
potentials at the lump positions, \ie $\{\delta\hat\varphi(\vec
x_l)\}$ and $\{\hat \Psi(\vec x_l)\}$ (and their gradients) as given
by Eqs.~(\ref{eq:phicalc}) and (\ref{eq:psicalc}). Finally, the mass
change is computed according to $\frac{dM_l}{dt} = -\beta_{l} M_l
\frac{d\hat\varphi}{dt}$. All these equations have mutual dependences
and can only be solved simultaneously. Cold dark matter, if included,
has to be treated with standard N-body techniques. The influence of
neutrino lumps on the matter component was studied in
\cite{Brouzakis10, Baldi11, Ayaita11}.

The aforementioned equations are only complete together with
functional relations $\beta_{l}(f_l, \hat \varphi)$ and $M_l(f_l,
\hat\varphi)$, \cf \fig{massfunctionbetas}, known at all times. As a
first approach, one may assume a time-independent relation. This is
reasonable if the lumps are virialized and hence their inner structure
is approximately frozen. We will explore the stability of individual
lumps in \sec{aspects}. Furthermore, the dependence on $\hat \varphi$
may be neglected if the derivative $\partial \beta/\partial \hat
\varphi$ or the variation of $\hat \varphi$ are sufficiently small. 

\begin{figure}[htb!]
	\begin{center}
		\psfrag{xlabel}[B][c][.8][0]{scale factor $a$}
		\psfrag{ylabel}[B][c][.8][0]{$\bar \rho_\nu\, a^3$
		[$10^{-9}\text{Mpc}^{-2}$]}
		\includegraphics[width=0.45\textwidth]{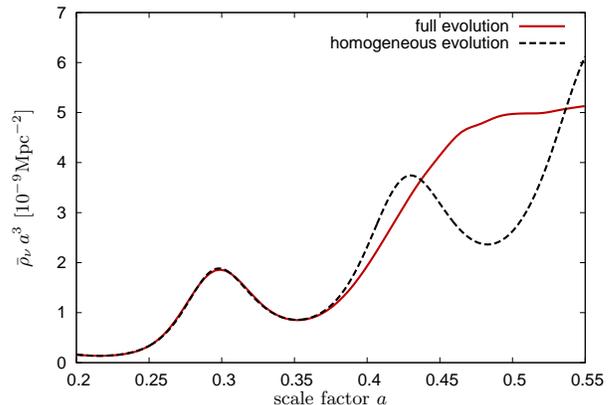}
	\end{center}
	\caption{Stabilization of the energy in neutrinos. The dashed line
	shows the evolution calculated by the background equations, the
	solid line is taken from a full simulation run.}
	\label{fig:rhoevolution}
\end{figure}

It is not clear whether $z \approx 1$ is late enough for the
virialization process to have sufficiently proceeded. A hint that the
cosmological configuration of the neutrinos is stabilizing, however,
is given by the evolution of the total neutrino energy $E_{(\nu)} =
\int d^3 x\, \sqrt{g^{(3)}}\,\rho_\nu(\vec x) \propto \bar \rho_\nu
a^3$, shown in \fig{rhoevolution}.
For $a \gtrsim 0.45$, one observes a transition to a regime with a
small constant slope. This would be compatible with a small monotonic
change of the large-scale cosmon field and an effective lump mass
depending on this field, corresponding to the expectation of
approximate mass freezing within neutrino lumps \cite{Nunes11}. This
may be taken as a hint that the neutrino lump fluid may become a
reasonable picture for $a \gtrsim 0.45$.

\section{Energy-momentum tensor of lumps}
\label{sec:energy}

In \sec{effective}, we had to assume properties of the energy-momentum
tensor associated with neutrino lumps. The derivations will be
provided in this section. An important result is the integrated
amplitude $A^{\mu\nu}$ of a single lump's energy-momentum tensor, see
\eq{amunu}. The derivation in \sec{single} includes the vanishing of
the total internal pressure in stable lumps. Next, in \sec{smoothed},
we will consider the term $\beta \hat T_\nu$, \cf \eq{smoothedsum},
which sources the energy-momentum exchange between cosmon and
neutrinos.

\subsection{Single lump}
\label{sec:single}

We now study a single neutrino lump described by its energy-momentum
tensor ${T^\mu}_\nu$ including contributions of the bound neutrinos
and the local cosmon field. The lump occupies a volume $V$; its
energy-momentum tensor vanishes outside. On scales much larger than
the lump size, it is useful to consider the amplitude 
\begin{equation}
	{A^\mu}_\nu = \int_{V}^{}d^3 x\, \sqrt{g^{(3)}}\,
	{T^\mu}_\nu.
	\label{eq:amunudef}
\end{equation}

We first switch to the rest frame of the lump where we will show
${A^\mu}_\nu = -M_l\,\delta^\mu_0\, \delta^0_\nu$. Let us therefor
consider the different components separately. Clearly, ${A^0}_0 = -
M_l$ by definition of the rest mass. $A^{i0} = P^i$ is the total
momentum and thus vanishes in the rest frame, whereby ${A^i}_0 = 0$.
It remains to show ${A^i}_j = 0$.

We assume that the lump is approximately static, \ie its
energy-momentum content in a physical volume is conserved,
\begin{equation}
	\partial_0 \left( a^3\,{T^\mu}_\nu \right) \approx 0,
	\label{eq:staticity}
\end{equation}
neglecting the metric perturbations. Together with the energy-momentum
conservation equation,
\begin{equation}
	0 = \nabla_\lambda {T^\lambda}_j = \partial_0 {T^0}_j + \partial_i
	{T^i}_j + 3 \frac{\dot a}{a} {T^0}_j,
	\label{eq:nabla}
\end{equation}
the staticity condition implies $\partial_i {T^i}_j = 0$. 

It is convenient to define the three-vector $\vec v = ({T^1}_j,
{T^2}_j, {T^3}_j)$ for a given column $j$. We have just shown
$\text{div}\, \vec v = 0$. From now on, we choose $i=j=1$ for
simplicity. The amplitude ${A^1}_1$ can then be written as
\begin{equation}
	{A^1}_1 = a^3 \int_{}^{} d x\,\int_{}^{}d y\,d z\ v_1 
	= a^3 \int_{}^{} d x\, \int_{S_x}^{} d \vec S \cdot \vec v,
	\label{eq:a11calc}
\end{equation}
where $S_x$ is the slice of $V$ normal to the $x$ direction. Outside
the lump, we extend the area $S_x$ to a closed surface. We can equally
integrate over this closed surface since there is no contribution
outside the lump. We conclude that the integral vanishes since
$\text{div}\,\vec v = 0$ inside the enclosed volume. This implies
${A^1}_1 = 0$. The derivation can equally be done for arbitrary $i$
and $j$, whereby ${A^i}_j = 0$.

In the presence of an external cosmon perturbation $\delta \hat
\varphi$ sourced by other lumps, the energy-momentum conservation used
in \eq{nabla} only applies to the full energy-momentum tensor
$T^{\mu\nu}_\text{tot}$ including the contribution due to $\delta \hat
\varphi$. Spatial variations of $\hat \varphi$ on the scale of the
lump are, however, small, such that $\partial_i T^i_{\text{tot}\,j}
\approx \partial_i {T^i}_j = \text{div}\, \vec v$. If the external
field $\delta \hat \varphi$ varies only slowly, the staticity
condition, \eq{staticity}, applies to the total energy-momentum tensor
as well.

The straightforward generalization of the rest-frame result gives the
amplitude ${A^\mu}_\nu = M_l\, u^\mu u_\nu/\gamma$ as anticipated in
\sec{description}. The lump, on scales larger than its size, is
described by a standard one-particle energy-momentum tensor
\begin{align}
	{T^\mu}_\nu &= \frac{1}{\sqrt{-g}}\int_{}^{}d\tau\,M_l\,u^\mu
	u_\nu\,\delta^{(4)}(x - x_l) \\
	&= \frac{1}{\sqrt{g^{(3)}}}\,\frac{M_l}{\gamma}\,u^\mu u_\nu\,
	\delta^{(3)}(\vec x - \vec x_l),
	\label{}
\end{align}
where we have used $u^0 = d x^0/d \tau$ and $\gamma =
\sqrt{-g_{00}}\,u^0$.

\subsection{Smoothed conservation equation}
\label{sec:smoothed}

In the effective description, the two dynamic components are the
collection of lumps (with the neutrino and a local cosmon
contribution) and the cosmon field $\hat \varphi$ outside the lumps,
which mediates the interaction. This differs from the usual split in
the neutrinos $T^{\mu\nu}_{(\nu)}$ and the cosmon
$T^{\mu\nu}_{(\varphi)}$ introduced in \sec{basics}. The total
energy-momentum content $T^{\mu\nu}_\text{tot}$ can thus be expressed
in two ways,
\begin{equation}
	T^{\mu\nu}_\text{tot} = T^{\mu\nu}_{(\nu)} + T^{\mu\nu}_{(\varphi)}
	= T^{\mu\nu}_\text{lumps} + T^{\mu\nu}_{(\hat\varphi)}.
	\label{}
\end{equation}
The neutrino contribution is completely contained in
$T^{\mu\nu}_\text{lumps}$. The cosmon field splits into $\varphi =
\hat \varphi + \delta \varphi_\text{loc}$, and the contribution of the
local perturbation $\delta \varphi_\text{loc}$ is attributed to the
energy-momentum tensor of the lumps. The part of the cosmon
energy-momentum tensor not depending on the local fluctuation
$\delta\varphi_\text{loc}$ is
\begin{equation}
	T^{\mu\nu}_{(\hat\varphi)} = 
	\partial^\mu\hat\varphi \,
	\partial^\nu\hat\varphi - g^{\mu\nu} \left( \frac{1}{2}
	\partial^\lambda \hat\varphi\, \partial_\lambda \hat\varphi +
	V(\hat\varphi) \right),
	\label{}
\end{equation}
which corresponds to the standard form of a scalar-field
energy-momentum tensor. 

Only the total energy-momentum tensor is conserved and we want to
investigate the energy-momentum flow between the components
$T^{\mu\nu}_{(\hat \varphi)}$ and $T^{\mu\nu}_\text{lumps}$. This will
yield an effective coupling $\beta_{l}$ between the lumps and $\hat
\varphi$. The four-divergence of $T^{\mu\nu}_{(\hat \varphi)}$ is
\begin{equation}
	\nabla_\lambda T^{\mu\lambda}_{(\hat \varphi)}
	= \left( \nabla^\lambda \nabla_\lambda \hat \varphi -
	V'(\hat\varphi) \right) \, \partial^\mu \hat \varphi.
	\label{eq:divthatphi}
\end{equation}
In order to evaluate the right-hand side, we employ the equation of
motion of the full cosmon field $\varphi$ inferred from
\eq{exchange1}:
\begin{equation}
	\nabla^\lambda \nabla_\lambda \varphi -
	V'(\varphi) = \beta T_{(\nu)}.
	\label{}
\end{equation}
Smoothing this relation at the scale $\lambda$ (\cf \sec{description})
at linear order in $\delta\varphi_\text{loc}$ and inserting into
\eq{divthatphi} yields
\begin{equation}
	\nabla_\lambda T^{\mu\lambda}_{(\hat\varphi)}
	= \beta \hat T_{(\nu)}\,\partial^\mu \hat \varphi.
	\label{eq:smoothcons1}
\end{equation}

The right-hand side can be expressed in terms of lump properties by
making use of the conservation equation for the total energy-momentum
tensor, $\nabla_\lambda T^{\mu\lambda}_{(\hat \varphi)} =
-\nabla_\lambda T^{\mu\lambda}_\text{lumps}$. The part
$\nabla_\lambda T^{\mu\lambda}_\text{lumps}$ can be analyzed in the
effective description where lumps are treated as point particles. The
equation of motion (\ref{eq:eom}) implies
\begin{equation}
	\nabla_\lambda T^{\mu\lambda}_\text{lumps}
	\approx -\sum_{\text{lumps }l}^{} \beta_{l} T_{l}
	\,\partial^\mu \hat \varphi.
	\label{eq:smoothcons2}
\end{equation}
Comparison with \eq{smoothcons1} yields
\begin{equation}
	\beta \hat T_{(\nu)}
	= \sum_{\text{lumps }l}^{} \beta_{l} \, \hat
	T_l,
	\label{}
\end{equation}
which is the relation used in \sec{evolution}.

\section{Aspects of stability}
\label{sec:aspects}

The effective description of the cosmological dynamics outlined in
\sec{effective} relies on the assumption of stable lumps. At the
current stage of the comprehensive simulation method \cite{Ayaita11},
however, it is not possible to track the evolution of lumps after $z
\approx 1$. In this section, we sketch some analytic arguments why
stable lumps are expected to form. We start with considerations
concerning the angular momentum, \sec{angular}, and construct an
explicit example of a static configuration using hydrodynamic
equations in \sec{hydrodynamic}. In the following, we neglect the
metric perturbations.

\subsection{Angular momentum}
\label{sec:angular}

The cosmon-mediated fifth force felt by the neutrinos is stronger
than but in some respects similar to gravity. The field equation for
$\delta\varphi$, \eq{kgpert}, can be compared to the usual
gravitational Poisson equation. The cosmon perturbation
$\delta\varphi$ thus plays the role of a potential --~similar to the
gravitational potential~-- in which a neutrino particle moves. In
contrast to the gravitational case, however, the particle changes its
mass $m_\nu = m_\nu(\varphi)$ while moving with velocity $u^\mu$
according to
\begin{equation}
	\dot m_\nu = -\beta m_\nu \, \frac{u^\lambda \partial_\lambda
	\varphi}{u^0}.
	\label{}
\end{equation}
The loss of mass when moving towards a minimum of the potential
implies, by momentum conservation, an additional acceleration
\cite{Ayaita11}. Hence, it has to be investigated whether neutrino
lumps are unstable, \ie continuously shrink to smaller sizes until
they are stabilized, \eg, by the degeneracy pressure
\cite{Brouzakis07}.

The cosmon-mediated fifth force, despite the mass variation along a
particle trajectory, shares an important property with gravity: the
conservation of angular momentum. For example, a single particle
moving in a spherically symmetric and static cosmon potential
$\varphi(r)$ (in physical coordinates) has the conserved angular
momentum
\begin{equation}
	L = \gamma\,m_\nu\, r^2 \dot\theta
	\label{}
\end{equation}
in polar coordinates $(r,\theta)$ and with the Lorentz factor
$\gamma$. The equation of motion, written for the radial momentum $p_r
= \gamma m_\nu \dot r$, then contains an angular momentum barrier,
which prevents the particle from falling into the center. It reads
\begin{equation}
	\dot p_r = \frac{L^2}{\gamma m_\nu r^3} +
	\frac{\beta m_\nu}{\gamma}\,\frac{d\varphi}{dr}.
\end{equation}
This is analogous to Newtonian gravity with an angular momentum
barrier $\propto L^2/r^3$ and an inward potential gradient. The only
difference is the variation of $m_\nu$ (and $\gamma$) along the
particle's trajectory. Since the mass decreases when approaching the
center, this even amplifies the angular momentum barrier.

Of course, these results for a test particle in a central potential
need not generalize to a distribution of particles forming a lump.
There, we define a neutrino angular momentum density
$l^{\mu\nu\alpha}_{(\nu)}$ as in special relativity, 
\begin{equation}
	l^{\mu\nu\alpha}_{(\nu)} = x^\mu T^{\nu\alpha}_{(\nu)}
	- x^\nu T^{\mu\alpha}_{(\nu)},
	\label{eq:lmunualpha}
\end{equation}
which, without the cosmon-neutrino coupling, would satisfy a
conservation equation $\nabla_\alpha (l^{i j\alpha}_{(\nu)}/a) = 0$
due to the conservation equation for $T^{\mu\nu}_{(\nu)}$. Here,
derivatives are taken with respect to comoving coordinates. Defining
the total spatial neutrino angular momentum
\begin{equation}
	L^{ij}_{(\nu)} \equiv \int_{}^{} d^3 x \, \sqrt{g^{(3)}}\,
	l^{ij0}_{(\nu)},
	\label{}
\end{equation}
the conservation equation for $l^{i j\alpha}_{(\nu)}$ in the
uncoupled case translates to the conservation law
\begin{equation}
	\partial_t \left( a^2 L^{ij}_{(\nu)} \right) = 0.
	\label{}
\end{equation}

With the coupling, \eq{exchange2}, we instead obtain
\begin{equation}
	\nabla_\alpha \left( a^{-1} l^{i j\alpha}_{(\nu)} \right) =
	-a^{-1}\beta T_{(\nu)}
	\left( x^i \partial^j \varphi - x^j \partial^i \varphi
	\right),
	\label{}
\end{equation}
and thus
\begin{align}
	\partial_t \left( a^2 L^{ij}_{(\nu)} \right) &= \partial_t
	\int_{}^{} d^3 x \, \sqrt{g^{(3)}}\,a^2 l^{ij0}_{(\nu)}
	\\
	&= \int d^3 x \,\sqrt{g^{(3)}}\, a^2 \beta T_{(\nu)} \left( x^i
	\partial^j \varphi - x^j \partial^i \varphi \right).
	\label{eq:partialta3l}
\end{align}
For a spherically symmetric lump and thus cosmon
potential $\varphi = \varphi(t,r)$, it is straightforward to show
\begin{equation}
	x^i \partial^j \varphi -
	x^j \partial^i \varphi
	= 0.
	\label{}
\end{equation}
In this case, the quantity $a^2 L^{ij}_{(\nu)}$ is indeed conserved.
This is related to the fact that a spherically symmetric scalar field
does not carry spatial angular momentum,
\begin{equation}
	l^{ij0}_{(\varphi)} = -\dot\varphi\,\left( x^i \partial^j \varphi
	- x^j \partial^i \varphi \right) = 0.
	\label{}
\end{equation}
The conservation of the total angular momentum $a^2 L^{ij}_\text{tot}$
then reduces to the conservation of $a^2 L^{ij}_{(\nu)}$.

Our considerations hold for an arbitrary isotropic and homogeneous
background metric. Thus, $a$ does not need to be the cosmic scale
factor but can also describe some local properties of the metric.
Fluctuations of the metric around the background metric as well as
fluctuations of the cosmon around an averaged field as $\hat \varphi$
in \eq{smoothedfield} can be added to the neutrino energy-momentum
tensor in \eq{lmunualpha}. The right-hand side of \eq{partialta3l}
involves then $\hat \varphi$ instead of $\varphi$ and $\beta_{l}$
instead of $\beta$, resulting in a reduction of the change of angular
momentum. We conclude that angular momentum conservation is similar to
standard gravity. This constitutes a strong hint for a dynamic
stabilization of the lump.

\subsection{Hydrodynamic balance}
\label{sec:hydrodynamic}

We will now study a neutrino lump within a hydrodynamic framework and
derive a balance equation for a simple class of lumps. For this
purpose, we will employ moments of the neutrino phase-space
distribution function $f(t, x^i, p_j)$ describing the distribution of
particles with comoving position $x^i$ and momentum $p_j = m_\nu u_j$.
A discussion of stability based on the Tolman-Oppenheimer-Volkoff
equation can be found in Ref.~\cite{Bernardini09}. For simplicity, we
will restrict ourselves to first-order relativistic corrections in
this section. The equations of motion for a neutrino particle under
the influence of the fifth force can then be written as
\begin{equation}
	\dot x^i = \frac{p^i}{m_\nu},\ \dot p_j = \left( 1 -
	\frac{p^k p_k}{2 m_\nu^2} \right) \beta m_\nu \partial_j
	\varphi.
	\label{eq:eomfirstorder}
\end{equation}
The fully relativistic
equation in terms of the four-velocity $u^\mu$ is presented in
\cite{Ayaita11}.

We will consider the following moments of the phase-space distribution
function $f$:
\begin{align}
	n &= \int_{}^{}d^3 p\, f(t, \vec x, \vec p),\label{eq:ndef}\\
	n U_i &= \int_{}^{} d^3 p\, \frac{p_i}{a m_\nu} f(t, \vec x, \vec
	p), \label{eq:Udef}\\
	\sigma_{ij} + n U_i U_j & = \int_{}^{}d^3 p\,
	\frac{p_i}{a m_\nu}\frac{p_j}{a m_\nu} f(t, \vec x, \vec p).
	\label{eq:sigmadef}
\end{align}
The quantities $n(t, \vec x)$, $\vec U(t, \vec x)$, and $\sigma_{ij}(t,
\vec x)$ are interpreted as the number
density, the locally averaged peculiar velocity, and the velocity
dispersion tensor, respectively. Their evolution equations can be
derived from the principle of particle conservation in phase-space,
which is expressed by the continuity equation
\begin{equation}
	\dot f + \frac{\partial (f \dot x^i)}{\partial x^i} +
	\frac{\partial( f \dot p_j)}{\partial p_j} = 0.
	\label{eq:continuityf}
\end{equation} 
The whole procedure is similar to the standard case of gravity (\cf
Ref.~\cite{Bernardeau01}) with the peculiarity of a varying mass
$m_\nu = m_\nu(\varphi)$.

Integrating over the momentum in \eq{continuityf} yields the zeroth
moment
\begin{equation}
	\dot n + \partial_{r_i} (n U_i) = 0,
	\label{eq:continuityn}
\end{equation}
with $\partial_{r_i} = a^{-1}\partial/\partial x^i$. A static number
density profile, $\dot n = 0$, is realized if the microscopic
motion adds locally up to zero, $\vec U = 0$. This is the case for
a locally isotropic velocity distribution. For this class of lumps, 
the equation for $\dot {\vec U}$, which follows by taking
the first moment of \eq{continuityf} and using the equations of motion
(\ref{eq:eomfirstorder}), takes
a particularly simple form:
\begin{equation}
	\dot U_i = - \frac{1}{n} \partial_{r_j}\sigma_{ij} + \beta
	\partial_{r_i} \varphi \left( 1 - \frac{3 \sigma}{2 n} \right) +
	\frac{1}{n} \sigma_{ij}\beta\partial_{r_j}\varphi,
	\label{eq:accelU}
\end{equation}
with $\sigma \equiv {\sigma^i}_i/3$.

For a static lump, we demand $\dot {\vec U} = 0$ in addition to $\dot
n = 0$. A glance at \eq{accelU} shows that this requires a certain
balance between the effective pressure $\propto \partial_{r_j}
\sigma_{ij}$, generated by the microscopic neutrino motion, and the
fifth force $\propto \beta \partial_{r_i} \varphi$. Assuming spherical
symmetry, the balance equation reads
\begin{equation}
	\frac{1}{n}\sigma' = \beta \varphi' \left(1 - \frac{\sigma}{2 n}
	\right),
	\label{eq:balance}
\end{equation}
with a prime denoting derivatives with respect to the radial
coordinate. Here, we have used $\sigma_{ij}=\sigma\, \delta_{ij}$.
Solving this equation together with the (radial) Klein-Gordon equation
for the cosmon field yields static lump configurations. These lump
configurations differ from the solutions discussed in
Ref.~\cite{Brouzakis07} since the stabilizing pressure is now provided
by the neutrino motion rather than by the degeneracy pressure.

\begin{figure}[htb!]
	\begin{center}
		\psfrag{xlabel}[B][c][.8][0]{physical radius $ar$
		[$h^{-1}$Mpc]}
		\psfrag{ylabel}[B][c][.8][0]{radial profile}
		\includegraphics[width=0.45\textwidth]{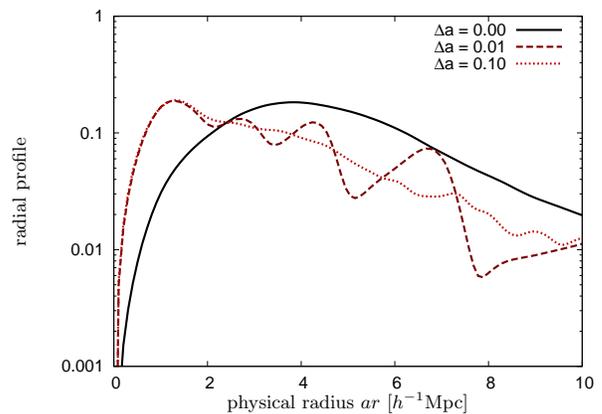}
	\end{center}
	\caption{The radial profile $\propto 4 \pi r^2\,n(r)$ of neutrinos
	inside the perturbed lump normalized to unity.}
	\label{fig:profile}
\end{figure}

At first sight, it is not clear whether this staticity condition
constitutes a stable equilibrium. We perform an exemplary numerical
check by simulating a single, isolated lump with the N-body technique
\cite{Ayaita11}. Rather than starting with a static lump configuration
by \eq{balance}, we use a somewhat smaller velocity dispersion
$\sigma$. In the subsequent evolution, the lump shrinks and the
neutrino pressure increases. Figure~\ref{fig:profile} shows how the
neutrino profile becomes more concentrated and indeed stabilizes. The
simulation evolves the lump in physical time $t$. For convenience, we
have translated time intervals $\Delta t$ to scale factor intervals
$\Delta a$ by the Hubble parameter at $z = 1$. 

\begin{figure}[htb!]
	\begin{center}
		\psfrag{xlabel}[B][c][.8][0]{$\Delta a$}
		\psfrag{ylabel}[B][c][.8][0]{integrated pressure $P(r)/M_l$}
		\includegraphics[width=0.45\textwidth]{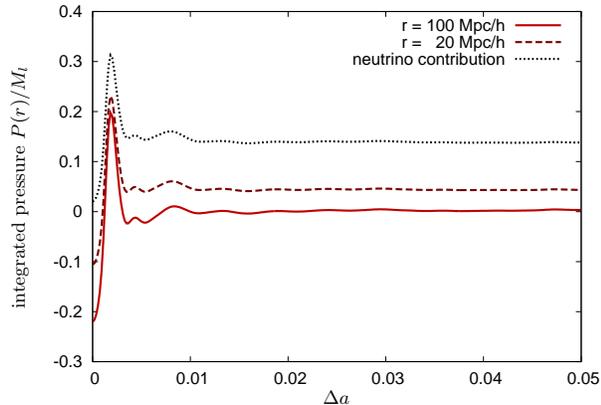}
	\end{center}
	\caption{The total pressure integrated to $r = 100\, h^{-1}$Mpc
	(red, solid) and $20\, h^{-1}$Mpc (dark red, dashed). For
	comparison, we also plot the neutrino contribution covering all
	neutrinos (black, dotted).}
	\label{fig:stabilization}
\end{figure}

The pressure cancellation between the contributions of neutrinos and
the cosmon perturbations, \cf \sec{lumps}, is established during the
stabilization process. This is shown in \fig{stabilization}.
Similar to \fig{cancellation}, we observe that the pressure
cancellation is only established at rather large distances from the
lump. At smaller distances, a residual positive pressure remains.

\section{Conclusion}
\label{sec:conclusion}

We have shown that a simplified, effective description of the
cosmological dynamics in the growing neutrino quintessence model is
possible. It bases upon describing stable cosmon-neutrino lumps as
nonrelativistic particles with an effective interaction. After the
main idea was given (\sec{lumps}), several aspects needed to be
investigated.

The first issue concerns the stability of the lumps. We have shown in
a hydrodynamic analysis of spherically symmetric lumps that the
neutrino velocity dispersion indeed stabilizes the lumps against the
attractive cosmon-mediated fifth force (\sec{hydrodynamic}). On more
general grounds, stability of the lumps is already expected by angular
momentum conservation which holds similarly to the gravitational case
(\sec{angular}). Stable lumps may then be characterized by the amount
of bound neutrinos. In numerical simulations of growing neutrino
quintessence, we have found lumps containing a fraction up to $\gtrsim
10^{-3}$ of all neutrinos in the Hubble volume, reaching a mass of
$\sim 10^{17}$ solar masses (\sec{description}). The total number of
identified lumps in the Hubble volume is of order $10^{4}$.

Second, it is not clear a priori that the lumps can be described as
particles. The most important aspect here is the vanishing of the
total internal pressure. The neutrinos, however, have reached high
velocities and an equation of state $w_\nu \approx 0.1$. We have shown
that --~under idealized conditions~-- the neutrino pressure is exactly
cancelled by a negative pressure contribution from the local cosmon
perturbations (\sec{energy}). A numerical check is given in
\fig{cancellation}. Under realistic conditions, the pressure
cancellation may not hold exactly but to a good approximation.
Approximate cancellation of neutrino and cosmon pressure occurs at a
characteristic radius $r_l$ that is substantially larger than the
radius of the neutrino core of the lump. For an effective particle
description, $r_l$ is the size of the lump. A fluid description
requires that the typical distance between lumps exceeds $r_l$.

Third and finally, a description of the cosmological dynamics requires
the equation of motion for the lumps and the field equation for the
smoothed field $\hat\varphi$ mediating the interaction between the
lumps. These equations have been derived in \sec{effective}. The
decisive quantity characterizing the lump interaction is the effective
cosmon-lump coupling $\beta_l$. For small lumps, it approaches the
fundamental coupling $\beta$ quantifying the cosmon-mediated fifth
force between neutrinos. For big lumps, the effective coupling
$\beta_l$ is suppressed by a factor of two to three as compared to
$\beta$. Since the attractive force is proportional to the squared
coupling, this corresponds to a suppression of the attraction by one
order of magnitude.

The effective description of growing neutrino quintessence complements
sophisticated numerical techniques as it provides physical insight
into the dynamics. Furthermore, the effective description could prove
useful in understanding the evolution for redshift $z < 1$, where
numerical simulations have not yet been successful \cite{Baldi11,
Ayaita11}. Quantitative results for low redshifts are needed to
eventually confront growing neutrino quintessence with observational
constraints.

The process of lump formation (in the redshift range $z \approx 1$ to
$2$) is very complex and still requires a thorough numerical
treatment. It constitutes, however, only a transitional period.
Thereafter, a physically sound and much simpler picture seems to
emerge.

The concepts and methods of averaging developed in this paper are
quite general for describing lumps in the presence of long-range
interactions mediated by a field. For our purpose, we needed a
relativistic treatment. A nonrelativistic version may be applied to
different clumping processes where separated interacting lumps are
forming. An example could be a gas or liquid of macromolecules for
which internal structure does not play a decisive role.

\begin{acknowledgments}
\end{acknowledgments}

We thank Marco Baldi for inspiring discussions and valuable ideas. We
acknowledge support from the DFG Transregional Collaborative Research
Centre on the ``Dark Universe.''

\bibliography{effective}

\end{document}